\documentclass[12pt,preprintnumbers,nofootinbib]{revtex4}
\usepackage{amsmath}
\usepackage{amsfonts}
\usepackage{amsbsy}
\usepackage{lscape} 
\usepackage{color}
\usepackage{graphicx,epsfig}
\usepackage[english]{babel}
\usepackage{latexsym}
\usepackage{amssymb}

\newcommand{\beq}{\begin{equation}}
\newcommand{\eeq}{\end{equation}}

\newcommand{\p}{\partial}
\newcommand{\wt}{\widetilde}
\newcommand{\ov}{\overline}
\newcommand{\md}{\mathcal{D}}

\newcommand{\suc}{{{\rm SU}_{\rm C}(3)}}
\newcommand{\sul}{{{\rm SU}_{\rm L}(2)}}
\newcommand{\ue}{{{\rm U}(1)}}
\newcommand{\el}{{\rm EM}}
\newcommand{\GeV}{{\rm GeV}}

\newcommand{\Leff}{{\mathcal{L}_{\rm eff}}}
\newcommand{\LQCD}{{\Lambda_{\rm QCD}}}
\newcommand{\uflavor}{\mathbf{1}_{\rm flavor}}
\newcommand{\CW}{C_2({\bf N_W})}
\newcommand{\CS}{C_2({\bf N_S})}
\newcommand{\lgr}{\left\lgroup}
\newcommand{\rgr}{\right\rgroup}

\begin{document}

\renewcommand{\thefootnote}{\fnsymbol{footnote}}


\title{Classification of Dimension 5 Lorentz Violating Interactions 
		in the Standard Model}

\author{Pavel A. Bolokhov$^{\,(a,b)}$ and Maxim Pospelov$^{\,(a,c)}$}

\affiliation{$^{\,(a)}${\it Department of Physics and Astronomy, University of Victoria, 
     Victoria, BC, V8P 1A1 Canada}\\
$^{\,(b)}${\it Theoretical Physics Department, St.Petersburg State University, Ulyanovskaya 1,
        Peterhof, St.Petersburg, 198504, Russia}\\
$^{\,(c)}${\it Perimeter Institute for Theoretical Physics, Waterloo,
Ontario N2J 2W9, Canada}
}

%
%

\begin{abstract}
	We give a complete classification of mass dimension five Lorentz-non-invariant interactions
    composed from the Standard Model fields, using the effective field theory approach. 
    We identify different classes of Lorentz violating operators, 
    some of which are protected against transmutation 
    to lower dimensions even at the loop level. Within each class of operators we determine  
    a typical experimental sensitivity to the size of Lorentz violation. 
\end{abstract}


\maketitle


%

\newpage

\section{Introduction}

	
	Lorentz symmetry is one of the most important ingredients of the 
	Standard Model (SM) of particles and fields, as well as its extensions at the 
electroweak scale. Even though there is a robust evidence that Lorentz symmetry is maintained to a 
high degree of accuracy, the searches of the so-called "Lorentz Violation" (LV) 
are still reasonably well-motivated. The intriguing possibility is that an a priori unknown physics at high 
energy scales could lead to a spontaneous breaking of Lorentz invariance by 
giving an expectation value to certain non-SM fields 
that carry Lorentz indices, such as vectors, tensors, and gradients of scalar fields
\cite{Kostelecky:1988zi}.
The interaction of these fields with  operators composed 
from the SM fields, fully Lorentz-symmetric before the spontaneous breaking, below the 
scale of the condensation will manifest itself as effective LV terms. Schematically, one 
has 
\begin{equation}
O^{\rm SM}_{\mu\nu...} C^{\mu\nu...} \to O^{\rm SM}_{\mu\nu...} \langle C^{\mu\nu...} \rangle,
\label{start}
\end{equation}
where $C^{\mu\nu...}$ is an external field that undergoes condensation, and 
$O^{\rm SM}_{\mu\nu...} $ is the SM operator that transforms under 
the Lorentz group. It would be fair to say that the dynamical breaking of Lorentz 
invariance is difficult to achieve in a consistent UV-complete theory, and most of the 
papers studying LV shy away from this issue. 
 Nevertheless, leaving aside the question of the dynamical LV, 
the spurion or the effective field theory approach to the problem 
\cite{Kost1,Colladay:1998fq} has been instrumental in 
comparing the sensitivity of various LV tests. 
	
If the generation of interaction \eqref{start} is a true UV phenomenon, it proves extremely useful 
to classify all operators of lowest dimensions coupled to a given spurion field $	C^{\mu\nu...}$.
At dimension three and four levels, this was done nearly a decade ago by Colladay and 
Kostelecky 
\cite{Kost1,Colladay:1998fq,CG}. 
Dimension three Lorentz violating interactions, {\em e.g.} $b^\mu \ov{\psi} \gamma_\mu\gamma_5\psi$ and 
 $a^\mu \ov{\psi} \gamma_\mu\psi$, must defy the naive dimensional counting, according to which $
 a^\mu$ and $b^\mu$ scale linearly with the UV energy scale responsible for LV, 
$a^\mu,b^\mu \sim \Lambda_{UV}$. 
This is, of course, in a stark contradiction with reality, 
where {\em e.g.} for the parameters $ b_\mu $ of light quarks the constraints of the order of $10^{-31}$ GeV 
can be derived from the limits on nucleon LV parameters
\cite{clock2}.
It can be hypothesized that spurions at dimension three and four levels are in fact 
not fundamental, but indeed effective, implying a different type of scaling, {\em e.g.} 
$a^\mu,b^\mu \sim \Lambda_{\rm IR}^2/\Lambda_{\rm UV}$, where $\Lambda_{IR}$ is the energy scale 
associated with the SM. The appearance of $\Lambda_{UV}$ in the denominator implies that above the 
scale $\Lambda_{\rm IR}$ one should be able to formulate all LV interactions in terms of the  
{\em higher-dimensional} LV operators.  

The purpose of this paper is to give a complete account of LV operators of dimension five 
in QED and in the SM. Previously, only certain subclasses of higher-dimensional operators 
have been considered. Notably, dimension six LV operators and 
their phenomenological consequences 
have been studied at length in the context of the canonical non-commutative field theories 
\cite{MPR1,Carroll:2001ws,Carlson:2001sw,UCSC}. 
The operators that introduce the UV modifications of the dispersion relations for particles 
have been a subject of intense investigations over a number of years 
\cite{Ted1,Vuc,MP:,Moore:2001bv,CG,GK,Bertolami:1996cq,Bolokhov:2006wu}. Finally, the 
supersymmetrized versions of the dimension five LV operators have been studied in detail in 
\cite{GrootNibbelink:2004za,Bolokhov:2005cj}. 
Our paper includes all previously known cases of LV dimension five interactions, but also 
extends  LV to operators that have never been discussed in the literature. 

Our intention is to write down a generic LV theory with mass dimension
	five interactions, compatible with the Standard Model and to classify
	the inherent LV operators. We count the dimension of the SM operators
	$O^{\rm SM}_{\mu\nu...}$, and thus dimension five corresponds to all 
	background LV spurions to be of the inverse energy scale. 
%
	It is easy to see that a theory with LV interactions of mass
	dimension five admits a more diverse set of operators and as a consequence 
    more LV backgrounds $C^{\mu\nu...}$
than one has at dimension three and four levels. Since there is a significant 
freedom in the choice of LV spurions, we take an approach that each of them represents 
an  irreducible Lorentz tensor structure, which leads to a significant facilitation of the 
analysis of loop effects, and often protects dimension five operators from transmuting into 
lower dimensions when the quantum effects are taken into account. 
	Thus, we augment the requirements specified in 
\cite{MP:} by demanding that LV spurions be irreducible tensors under the Lorentz group transformations.
	In total, these  conditions look as follows: 
	an LV operator of specific dimension must be
\begin{itemize}
	\item gauge invariant
	\item Lorentz invariant, after contraction with a background tensor
	\item not reducible to lower dimension operators by the equations
		of motion
	\item not reducible to a total derivative
	\item coupled to an irreducible background tensor.
\end{itemize}

	We find that operators built in this manner can be 
	subdivided into three main groups.
	The first group includes the ``unprotected operators'', {\it i.e.} those which can
	generate lower-dimensional interactions by developing quadratic divergencies.
	Such operators are therefore dangerous, and as a rule,  severely constrained
	by strong limits on lower dimensional operators multiplied
	by the square of the UV scale.
	The second group is the UV-enhanced operators, which induce modifications of  
the dispersion relations that grow with particle's energy. These operators induce new 
testable LV signatures in the laboratory experiments and in astrophysics, and are severely 
constrained by both.
	The last group is formed by ``soft LV interactions'' which are protected from 
    developing quadratic divergencies at loop level and do not significantly modify the 
	propagation of energetic particles in the UV.  Typically, such operators are constrained 
     by the laboratory searches of spatial anisotropy.



	The structure of this paper is as follows. In Section II we analyze the case of 
 Quantum Electrodynamics (QED), extended by all possible 
    LV interactions  of dimension five.
QED is one of the most popular testing grounds for LV
\cite{CFJ,CG,Ted1,MP:,Bolokhov:2005cj}, and the
	detailed study of LV QED facilitates a smooth transition to the Standard Model. 
	Within QED, we develop one-loop renormalization group equations for the LV interactions. 
	Going over to the Standard Model in Section III, we observe that the chirality of matter fields 
    imposes further restrictions on the type of admissible LV interactions. 
	However, the abundance of field content makes possible for more diverse 
	structures and links between them.
	A complete RG analysis in the Standard Model may be highly desirable for 
	refining the phenomenological constraints on LV operators. However, due to the excessively 
	complicated structure of interactions, we only elaborate on the example of 
	operators which modify dispersion relations, which are of the most phenomenological interest.
	In Section IV, we subdivide operators into the major groups according to their phenomenological 
consequences and give a brief account of typical limits one can expect from the currently available 
tests of Lorentz violation.


\section{Dimension 5 Operators in QED}

	In order to build a Lorentz-violating extension of QED, we take an approach
	similar to Myers-Pospelov electrodynamics
\cite{MP:}.
	The Lagrangian of QED was modified by adding
	a number of LV operators which were generated by an absolutely
	symmetric 3-rank irreducible tensor background.
	Originally, the choice of symmetric tensors was motivated by the fact that
	LV operators can modify the dispersion relations, and also that
	they do not induce dangerous quadratic divergencies.
Our intention is to classify {\it all} dimension five operators in 
	Quantum Electrodynamics, and thus the list of the external LV tensors will 
necessarily be expanded. 
	Generic operators will produce new non-minimal interactions between
	the electron and the photon. 
	The LV extension of the photon sector of QED appears to be the most simple, whereas
	the matter sector shows a rich structure of LV terms.

\subsection{Purely Gauge Operators in QED}

	Dimension five LV interactions can admit LV backgrounds up to rank five.
	Higher ranks can appear only in combination with operators of dimension six or higher.
	There are 26 numbered Young tableaux to consider, which in fact 
	lead to only one LV operator.

	It can be shown that a generic content of a gauge
	invariant tensor has to be bilinear in the field strength  $ F_{\mu\nu} $ 
     and contain one extra derivative, 
	which must be a covariant derivative in the case of a non-abelian field.


	The only non-vanishing terms that satisfy these properties are
\begin{equation}
\label{generic_gauge}
	F_{\mu\nu} \partial^\nu \widetilde{F}^{\mu\rho}~,\qquad
	F_{\mu\nu} \partial^\nu F^{\rho\sigma}~,\qquad
	F_{\mu\lambda} \partial_\nu \widetilde{F}^{\rho\lambda}\qquad
	\text{and}\qquad
	F^{\mu\nu} \partial^\lambda F^{\rho\sigma}~.
\end{equation}
	It can be easily seen that the first two terms are {\it reducible}
	on the equations of motion, and, in accord with our requirements
	should be ignored. 
	Amongst the two structures left, 
$ F_{\mu\lambda} \p_\nu \widetilde{F}_\rho^{\phantom{\rho}\lambda} $ and
$ F^{\mu\nu} \partial^\lambda F^{\rho\sigma} $,
	the first has been studied in 
\cite{MP:} 
	and shown to modify the dispersion relations of the photon. 
	It was shown in particular that this operator has to be 
	contracted with an
	irreducible absolutely symmetric tensor,
\begin{equation}
\label{QED_gauge}
	C^{\mu\nu\rho}\, F_{\mu\lambda} \p_\nu 
	\wt{F}_\rho^{\phantom{\rho}\lambda}~,
	\qquad C^{\mu\phantom{\mu}\rho}_{\phantom{\mu}\mu} = 0~.
\end{equation}
	 Conditions of absolute symmetry and irreducibility of
	the tensor $ C^{\mu\nu\rho} $ follow from the requirement of
	independence of this operator of the lower-rank operators
	of \eqref{generic_gauge}, which is also a way of protection against
the  mixing with such operators at the loop level.  

	The last structure in (\ref{generic_gauge}), the five-index object
$ F^{\mu\nu} \partial^\lambda F^{\rho\sigma} $,
	upon a naive substitution
	into the equations of motion, 
	seems to modify the dispersion relations in a manner similar to \eqref{QED_gauge}.
	However, that would be a misleading conclusion.
	As in the case of the 3-rd rank operator just discussed,
one needs to separate it from all 
	lower-rank interactions.
	In other words, one needs to subtract all possible 
	$ g^{\mu\nu} $ and $ \epsilon^{\mu\nu\rho\sigma} $ 
	traces of this term, and {\it then} substitute it into the
	equations of motion.
	It turns out that this operator is completely expressible
	in terms of its $ \epsilon^{\mu\nu\rho\sigma} $-trace, which
	coincides with the operator $ C^{\mu\nu\rho} $:
\begin{eqnarray*}
\lefteqn{
	F_{\mu\nu} \p_\lambda F_{\rho\sigma} = 
	} &&
	\\
	&&
	-~~ 
	\frac{1}{5}\,\epsilon_{\mu\nu\rho\chi} \,
	\wt{F}^{\zeta\chi}\p_\lambda F_{\zeta\sigma} 
 	~~+~~
	\frac{1}{5}\,\epsilon_{\mu\nu\sigma\chi} \,
	\wt{F}^{\zeta\chi}\p_\lambda F_{\zeta\rho} 
	~~+~~
	\frac{1}{5}\,\epsilon_{\rho\sigma\mu\chi} \,
	\wt{F}^{\zeta\chi}\p_\lambda F_{\zeta\nu} 
 	~~-~~
	\frac{1}{5}\,\epsilon_{\rho\sigma\nu\chi} \,
	\wt{F}^{\zeta\chi}\p_\lambda F_{\zeta\mu}
	\\
	&&
	-~~
	\frac{1}{10}\,\epsilon_{\mu\lambda\rho\chi} 
	\wt{F}^{\zeta\chi}\p_\nu F_{\zeta\sigma} 
	~~+~~
	\frac{1}{10}\,\epsilon_{\nu\lambda\rho\chi} 
	\wt{F}^{\zeta\chi}\p_\mu F_{\zeta\sigma} 
	~~+~~
	\\
	&&
	+~~
	\frac{1}{10}\,\epsilon_{\mu\lambda\sigma\chi} 
	\wt{F}^{\zeta\chi}\p_\nu F_{\zeta\rho} 
	~~-~~
	\frac{1}{10}\,\epsilon_{\nu\lambda\sigma\chi} 
	\wt{F}^{\zeta\chi}\p_\mu F_{\zeta\rho}
	~.
\end{eqnarray*}
	This relation shows that it is not possible to bring the rank five operator
	to an irreducible form,  and consequently there is no dimension 5 LV interaction 
    contracted with an irreducible rank five tensor. 
We conclude that the only possible
	LV operator in QED is $ C^{\mu\nu\rho} $.  
	All these arguments straightforwardly extend to a non-abelian gauge
	field.

\subsection{Matter Sector of QED}

	In contrast to what we have seen in the gauge sector, 
	the LV terms in the matter sector of QED have much
	wider variety.
	The reason for that is that the operators can be formed both
	by using covariant derivatives $ \cal{D}_\mu $ and by 
	inserting gamma matrices.
	
	In order to make the enumeration of operators more systematic, we use 
     Young tableaux, see Appendix \ref{young}.
	Omitting the details, we show the result for the LV operators
	in the matter sector, 
\begin{eqnarray}
\nonumber
\lefteqn{
	\mathcal{L}_{\rm QED}^{\rm matter} = 
	}
	\\
\nonumber
	&&
	\phantom{+~~}
	[c_1^{\mu}\cdot
	\ov{\psi}\, \gamma^\lambda F_{\mu\lambda} \psi {}^{+}] ~~+~~
	[c_2^{\mu}\cdot
	\ov{\psi}\, \gamma^\lambda \gamma^5 F_{\mu\lambda} \psi {}^{-}] ~~+~~
	\wt{c}_1^{\mu}\cdot
	\ov{\psi}\, \gamma^\lambda \wt{F}_{\mu\lambda} \psi {}^{+} ~~+~~
	\wt{c}_2^{\mu}\cdot
	\ov{\psi}\, \gamma^\lambda \gamma^5 \wt{F}_{\mu\lambda} \psi {}^{-}
	\\
\nonumber
	&&
	+~~
	f_1^{\mu\nu}\cdot
	\ov{\psi}\, F_{\mu\nu} \psi {}^{-} ~~+~~
	f_2^{\mu\nu}\cdot
	\ov{\psi}\, F_{\mu\nu} \gamma^5 \psi {}^{-} ~~+~~
	h_1^{\mu\nu}\cdot
	\ov{\psi}\, \mathcal{D}_{(\mu} \mathcal{D}_{\nu)} \psi {}^{+} ~~+~~
	h_2^{\mu\nu}\cdot
	\ov{\psi}\, \mathcal{D}_{(\mu} \mathcal{D}_{\nu)} \gamma^5 \psi {}^{+} 
	\\
\label{QED_matter}
	&&
	+~~
	C_1^{\mu\nu\rho}\cdot
	\ov{\psi}\, \gamma_{(\mu} 
		\mathcal{D}_\nu \mathcal{D}_{\rho)} \psi {}^{-} ~~+~~
	C_2^{\mu\nu\rho}\cdot
	\ov{\psi}\, \gamma_{(\mu} \gamma^5
	\mathcal{D}_\nu \mathcal{D}_{\rho)} \psi {}^{+} 
	\\
\nonumber
	&&
	+~~
	D_1^{\mu\nu\rho}\cdot
	\ov{\psi}\, \gamma_{(\mu} F_{\rho)\nu} \psi {}^{+} ~~+~~
	D_2^{\mu\nu\rho}\cdot
	\ov{\psi}\, \gamma_{(\mu} F_{\rho)\nu} \gamma^5 \psi {}^{-} 
	\\
\nonumber
	&&
	+~~
	E_1^{\mu\nu\rho\lambda}\cdot
	\ov{\psi}\, \sigma_{\mu)\nu} \mathcal{D}_{(\rho}\mathcal{D}_\lambda 
			\psi {}^{-} ~~+~~
	E_2^{\mu\nu\rho\lambda}\cdot
	\ov{\psi}\, \sigma_{\mu)(\lambda} F_{\rho)(\nu} \psi {}^{+} ~~+~~
	E_3^{\mu\nu\rho\lambda}\cdot
	\ov{\psi}\, \sigma_{\mu)[\nu} F_{\rho](\lambda} \psi {}^{+} 
	\\
\nonumber
	&&
	+~~
	E_4^{\mu\nu\rho\lambda}\cdot
	\ov{\psi}\, \left ( 
		\sigma_{\mu)[\nu} \mathcal{D}_{\rho]}\mathcal{D}_{(\lambda}
		~-~
		\sigma_{\nu](\mu} \mathcal{D}_{\lambda)}\mathcal{D}_{[\rho}
		~+~
		2\, \sigma_{\nu\rho} \md_{(\mu} \md_{\lambda)}
		\right ) \psi {}^{-} 
	~.
\end{eqnarray}

	In this formula,  $+$ and $-$ superscripts refer to
	the parity of the corresponding LV term under the charge conjugation.
	We stress again that all structures shown here assume
	their coefficients to be  irreducible tensors of the 
	corresponding rank. 
	Square brackets over the first two operators, $ c_1^\mu $ and $ c_2^\mu $, 
indicate that these two terms vanish upon the use of EOM, but we list them 
	for the reason they become nontrivial in the non-abelian case.

	We would like to make a side note on the symmetrizations in the 
	interactions in \eqref{QED_matter} and in subsequent formulae.
	We take the field operators to have certain symmetries 
	(dictated by the corresponding Young tableaux), while their Wilson
	coefficients to be just traceless tensors. 
	Equivalently, one could have cast all symmetrizations onto the
	Wilson coefficients, {\it e.g.}
$	E_1^{\mu\nu\rho\lambda}
	\ov{\psi}\, \sigma_{\mu)\nu} \mathcal{D}_{(\rho}\mathcal{D}_\lambda 
	\psi 
	\;\to\;
	E_1^{\mu)\nu(\rho\lambda}
	\ov{\psi}\, \sigma_{\mu\nu} \mathcal{D}_{\rho}\mathcal{D}_\lambda 
	\psi $,
	or just imply $ E_1^{\mu\nu\rho\lambda} $ to obey the corresponding
	symmetries: 
$	E_1^{\mu\nu\rho\lambda}\cdot
	\ov{\psi}\, \sigma_{\mu\nu} \mathcal{D}_{\rho}\mathcal{D}_\lambda 
	\psi $.
	We emphasize that this is only a matter of notation, and choose
	to expose the symmetry properties of tensors via explicit 
	symmetrizations on the Lorentz indices of the field operators.


\subsection{1-loop RG coefficients}

	If the violation of Lorentz invariance is a true UV phenomenon, one 
has to evaluate the operators down to the IR scale, where the 
majority of tests is performed. For this purpose, we study
	the renormalization group (RG) equations for operators
	\eqref{QED_gauge} and \eqref{QED_matter}.
	The RG running brings about the change in 
    the magnitude of Wilson coefficients 
    and mixing of different operators.

	Due to a rather large number of LV operators, one might expect that 
    this  mixings can be rather complicated.
	However, two reasons, namely the discrete symmetries and irreducibility
	of the Lorentz tensors, reduce this mixing to
	a minimum.
	The charge conjugation symmetry, which is an exact symmetry in QED,
	prevents the mixing of  $C$-odd and $C$-even operators.
	The irreducibility of the background tensors dictates that any tensor of
	higher rank will not mix with a tensor of a lower rank.
	Thus, only operators of the same rank can admix to each other.
	
	A brief examination of \eqref{QED_matter} reveals that
	$ \wt{c}_1^\mu $ cannot mix with $ \wt{c}_2^\mu $ due to $C$-parity.
	Similarly, $ f_{1,2}^{\mu\nu} $ cannot admix to $ h_{1,2}^{\mu\nu} $,
	but $ f^{\mu\nu}_1 \leftrightarrow f^{\mu\nu}_2 $ and
	$ h^{\mu\nu}_1 \leftrightarrow h^{\mu\nu}_2 $ mixings are allowed.
	At the level of rank three tensors, the photon
	operator \eqref{QED_gauge} is even under charge conjugation and
	therefore it can  mix only with the $ C_2^{\mu\nu\rho} $ operator.


	As we have admitted generic tensor structures to the theory
	we need to ensure that the latter is free of quadratic divergencies.
	It is obvious that quadratically divergent operators must necessarily
	couple to a vector background, as there are no dimension
	three structures which would be $CPT$-odd and contracted with a tensor
	background simultaneously. 
	In our list \eqref{QED_matter}, only the $ \wt{c}_1^\mu $ term generates 
quadratically divergent corrections to LV dimension three operators. The result 
of explicit computation gives the following set of RG equations:

\begin{align*}
	&
	\frac{d} {d t}\, \wt{c}_1^\mu ~\overset{\text{formally}}{=}~
		-\, \frac{13e^2}{96\pi^2}\, \wt{c}_1^\mu
	&&
	\frac{d} {d t}\, \wt{c}_2^\mu ~=~
		\frac{e^2}{32\pi^2}\, \wt{c}_2^\mu
	\\
	&
	\frac{d} {d t}\, f_{1,2}^{\mu\nu} ~=~
		-\, \frac{7e^2}{24\pi^2}\, f_{1,2}^{\mu\nu}
		~+~
		i\, \frac{7e^2}{48\pi^2}\, \wt{f}{}_{2,1}^{\mu\nu}
	&&
	\frac{d} {d t}\, h_{1,2}^{\mu\nu} ~=~
		\frac{e^2}{6\pi^2}\ h_{1,2}^{\mu\nu} 
	\\
	&
	\frac{d} {d t}\, C_1^{\mu\nu\rho} ~=~
		\frac{25e^2}{48\pi^2}\, C_1^{\mu\nu\rho}
	&&
	\frac{d} {d t}\, C_2^{\mu\nu\rho} ~=~
		\frac{25e^2}{48\pi^2}\, C_2^{\mu\nu\rho}
		~-~
		\frac{5e^2}{48\pi^2}\, C^{\mu\nu\rho}
	\\
	&&&
	\frac{d} {d t}\, C^{\mu\nu\rho} ~=~
		\frac{e}{48\pi^2}\, C_2^{\mu\nu\rho} 
		~-~
		\frac{e^2}{6\pi^2}\, C^{\mu\nu\rho}
	\\
	&
	\frac{d} {d t}\, D_1^{\mu\nu\rho} ~=~
		-\, \frac{e^2}{16\pi^2}\, D_1^{\mu\nu\rho}
	&&
	\frac{d} {d t}\, D_2^{\mu\nu\rho} ~=~
		\frac{5e^2}{48\pi^2}\, D_2^{\mu\nu\rho}
	\\
	&
	\frac{d} {d t}\, E_1^{\kappa\mu\nu\rho} ~=~
		\frac{13e^2}{24\pi^2}\, E_1^{\kappa\mu\nu\rho}
	&&
	\frac{d} {d t}\, E_3^{\kappa\mu\nu\rho} ~=~
		\frac{e^2}{3\pi^2}\, E_3^{\kappa\mu\nu\rho}
	\\
	&
	\frac{d} {d t}\, E_2^{\kappa\mu\nu\rho} ~=~
		\frac{e^2}{12\pi^2}\, E_2^{\kappa\mu\nu\rho}
	&&
	\frac{d} {d t}\, E_4^{\kappa\mu\nu\rho} ~=~
		\frac{e^2}{8\pi^2}\, E_4^{\kappa\mu\nu\rho}
	~.
\end{align*}	
	Here we have introduced $ \wt{f}{}_{1,2}^{\mu\nu} \,\equiv\, \frac{1}{2} 
		\epsilon^{\mu\nu\rho\sigma} (f_{1,2})_{\rho\sigma} $.
	As anticipated, most operators renormalize independently.
	It is also clear that one can easily form the linear combinations
	of LV interactions that are eigenvectors of one-loop RG equations.

\section{Classification of Operators of Dimension V in the Standard Model}

	In the Standard Model, the set of LV operators is more complicate,
	due to the wider gauge group. 
	Since the LV physics in our approach is associated with the UV scale, 
    the LV operators must 
	respect all the symmetries which are present at that scale.
	Although the UV physics and its symmetries are not known, it is quite natural 
    to require that LV interactions be invariant under 
	$ \suc\, \otimes $ $ \sul\, \otimes $ $ \ue $.
	Clearly the existence of families causes coefficients
	of all LV interactions in the matter sector \eqref{QED_matter} to be 
	matrices in the flavor space
\cite{Colladay:1998fq}.
	Furthermore, the presence of the Higgs sector creates new
	possibilities for LV interactions. 
	However, there is one simplification arising from intrinsic chirality of
	SM spinors, which together with gauge invariance 
   would essentially prohibit all $E^{\kappa\mu\nu\rho}$ operators at dimension five level.
In the rest of this section, we present our results for LV operators in different 
sectors of the SM. 


\subsection{Operators in the Gauge Sector of the Standard Model}

	As in the QED case, the gauge sector is the simplest since we already know
	that the only possible LV gauge structure is 
	\eqref{QED_gauge}.
	Thus we replicate this structure for the three gauge groups of the SM:
\begin{equation}
\label{SM_gauge}
	\mathcal{L}_{\rm SM}^{\rm gauge} = 
	C_{\ue}^{\mu\nu\rho} \cdot 
	F_{\mu\lambda} \,\partial_\nu\, \wt{F}_\rho^{\phantom{\rho}\lambda}
	~~+~~
	C_{\sul}^{\mu\nu\rho} \cdot 
	{\rm tr}\,
	W_{\mu\lambda} \,\mathcal{D}_\nu\, \wt{W}_\rho^{\phantom{\rho}\lambda}
	~~+~~
	C_{\suc}^{\mu\nu\rho} \cdot 
	{\rm tr}\,
	G_{\mu\lambda} \,\mathcal{D}_\nu\, \wt{G}_\rho^{\phantom{\rho}\lambda}
	~.
\end{equation}


\subsection{Matter Sector of the Standard Model}
\label{matter_SM}
	Although the matter sector of the Standard Model is more diverse
	than that of QED, the number of ``types'' of operators is smaller.
	Due to chirality of both leptons and quarks, the structures 
	with an even number of $ \gamma $-matrices in \eqref{QED_matter}
	are not $ \sul $-gauge invariant. 
	That greatly simplifies the structure of the LV Lagrangian,
	as one can only have operators with an odd number of gamma
	matrices.
	In the resulting Lagrangian we have to abandon the $C$-parity eigenstates, 
    \eqref{QED_matter} and list operators using $V-A$ and $V+A$ combinations of Dirac matrices. 
	
	Since in QED
\begin{equation}
\label{Fmn}
	\mathcal{D}_{[\mu}\mathcal{D}_{\nu]} = i e F_{\mu\nu}~, 
\end{equation} 
	the first two terms in \eqref{QED_matter} actually vanish on 
	the equations of motion.
	However, with the exception of right-handed leptons, 
	the covariant derivatives for SM field contain different gauge potentials. 
	For example, for quarks one has
\begin{equation}
\label{FmnQCD}
	\mathcal{D}_{[\mu}\mathcal{D}_{\nu]} = 
	i\, Y g'\, F_{\mu\nu} ~+~ 
	i\, g\, W_{\mu\nu} ~+~
	i\, g_3\, G_{\mu\nu}~,
\end{equation}
	where $ Y $ is the hypercharge of the quark.
	The use of equations of motion allows then to express one of the operators 
$ \ov{Q}\, \gamma^\lambda F_{\mu\lambda} Q~ $,
$ \ov{Q}\, \gamma^\lambda W_{\mu\lambda} Q~ $ and
$ \ov{Q}\, \gamma^\lambda G_{\mu\lambda} Q $
	in terms of the other two but one cannot eliminate such operators completely.
	Taking this into account, 
	in the quark sector 
	one obtains the following LV interactions:
\begin{eqnarray}
\nonumber
\lefteqn{
	\mathcal{L}_{\rm SM}^{\rm quark} =
	}
	\\
\nonumber
	&&
	\phantom{+~~}
	c_{Q,1}^\mu \cdot
	\ov{Q}\, \gamma^\lambda F_{\mu\lambda}\, Q 
	~~+~~
	c_{Q,3}^\mu \cdot
	\ov{Q}\, \gamma^\lambda W_{\mu\lambda}\, Q 
	~~+~~
	c_{u}^\mu \cdot
	\ov{u}\, \gamma^\lambda F_{\mu\lambda}\, u
	~~+~~ 
	c_{d}^\mu \cdot
	\ov{d}\, \gamma^\lambda F_{\mu\lambda}\, d
	~~+
	\\
\label{SM_quark}
	&&
	+~~
	\wt{c}_{Q,1}^\mu \cdot
	\ov{Q}\, \gamma^\lambda \wt{F}_{\mu\lambda}\, Q 
	~~+~~
	\wt{c}_{Q,2}^\mu \cdot
	\ov{Q}\, \gamma^\lambda \wt{W}_{\mu\lambda}\, Q 
	~~+~~
	\wt{c}_{Q,3}^\mu \cdot
	\ov{Q}\, \gamma^\lambda \wt{G}_{\mu\lambda}\, Q 
	~~+
	\\
\nonumber
	&&
	+~~
	\wt{c}_{u,1}^\mu \cdot
	\ov{u}\, \gamma^\lambda \wt{F}_{\mu\lambda}\, u
	~~+~~ 
	\wt{c}_{u,3}^\mu \cdot
	\ov{u}\, \gamma^\lambda \wt{G}_{\mu\lambda}\, u
	~~+~~ 
	\wt{c}_{d,1}^\mu \cdot
	\ov{d}\, \gamma^\lambda \wt{F}_{\mu\lambda}\, d
	~~+~~ 
	\wt{c}_{d,3}^\mu \cdot
	\ov{d}\, \gamma^\lambda \wt{G}_{\mu\lambda}\, d
	~~+
	\\
\nonumber
	&&
	+~~
	C_Q^{\mu\nu\rho} \cdot
	\ov{Q}\, \gamma_{(\mu} \md_\nu \md_{\rho)}\, Q
	~~+~~
	C_u^{\mu\nu\rho} \cdot
	\ov{u}\, \gamma_{(\mu} \md_\nu \md_{\rho)}\, u
	~~+~~
	C_d^{\mu\nu\rho} \cdot
	\ov{d}\, \gamma_{(\mu} \md_\nu \md_{\rho)}\, d
	~~+
	\\
\nonumber
	&&
	+~~
	D_{Q,1}^{\mu\nu\rho} \cdot
	\ov{Q}\, \gamma_{(\mu} F_{\rho)\nu}\, Q
	~~+~~
	D_{Q,2}^{\mu\nu\rho} \cdot
	\ov{Q}\, \gamma_{(\mu} W_{\rho)\nu}\, Q
	~~+~~
	D_{Q,3}^{\mu\nu\rho} \cdot
	\ov{Q}\, \gamma_{(\mu} G_{\rho)\nu}\, Q
	~~+
	\\
\nonumber
	&&
	+~~
	D_{u,1}^{\mu\nu\rho} \cdot
	\ov{u}\, \gamma_{(\mu} F_{\rho)\nu}\, u
	~~+~~
	D_{u,3}^{\mu\nu\rho} \cdot
	\ov{u}\, \gamma_{(\mu} G_{\rho)\nu}\, u
	~~+
	\\
\notag
	&&
	+~~
	D_{d,1}^{\mu\nu\rho} \cdot
	\ov{d}\, \gamma_{(\mu} F_{\rho)\nu}\, d
	~~+~~
	D_{d,3}^{\mu\nu\rho} \cdot
	\ov{d}\, \gamma_{(\mu} G_{\rho)\nu}\, d
	~.
\end{eqnarray}
	Here all coefficients are assumed to be Hermitian matrices in the 
	flavor space, {\it e.g.}
\[
	c_{Q,1}^\mu \cdot
	\ov{Q}\, \gamma^\lambda F_{\mu\lambda}\, Q 
	\equiv
	\left(c_{Q,1}^\mu\right)_{ik} \cdot
	\ov{Q}{}_i\, \gamma^\lambda F_{\mu\lambda}\, Q_k
	~. 
\]



	Similarly, LV interactions in the lepton sector of the Standard Model
	take the form:
\begin{eqnarray}
\nonumber
\lefteqn{
	\mathcal{L}_{\rm SM}^{\rm lepton} =
	}
	\\
\nonumber
	&&
	\phantom{+~~}
	c_L^\mu \cdot
	\ov{L}\, \gamma^{\lambda} F_{\mu\lambda}\, L
	~~+~~
	\wt{c}_{L,1}^\mu \cdot
	\ov{L}\, \gamma^{\lambda} \wt{F}_{\mu\lambda}\, L
	~~+~~
	\wt{c}_{L,2}^\mu \cdot
	\ov{L}\, \gamma^{\lambda} \wt{W}_{\mu\lambda}\, L
	+~~
	\\
\nonumber
	&&
	+~~
	\wt{c}_{\nu}^\mu \cdot
	\ov{\psi}{}_\nu\, \gamma^{\lambda} \wt{F}_{\mu\lambda}\, \psi_\nu
	~~+~~
	\wt{c}_{e}^\mu \cdot
	\ov{\psi}{}_e\, \gamma^{\lambda} \wt{F}_{\mu\lambda}\, \psi_e
	~~+~~
	\\
\nonumber
	&&
	+~~
	C_L^{\mu\nu\rho} \cdot
	\ov{L}\, \gamma_{(\mu} \md_\nu \md_{\rho)}\, L
	~~+~~
	C_{\nu}^{\mu\nu\rho} \cdot
	\ov{\psi}{}_\nu\, \gamma_{(\mu} \md_\nu \md_{\rho)}\, \psi_\nu
	~~+~~
	C_{e}^{\mu\nu\rho} \cdot
	\ov{\psi}{}_e\, \gamma_{(\mu} \md_\nu \md_{\rho)}\, \psi_e
	~~+
	\\
\label{SM_lepton}
	&&
	+~~
	D_{L,1}^{\mu\nu\rho} \cdot
	\ov{L}\, \gamma_{(\mu} F_{\rho)\nu}\, L
	~~+~~
	D_{L,2}^{\mu\nu\rho} \cdot
	\ov{L}\, \gamma_{(\mu} W_{\rho)\nu}\, L
	~~+
	\\
\nonumber
	&&
	+~~
	D_{\nu}^{\mu\nu\rho} \cdot
	\ov{\psi}{}_\nu\, \gamma_{(\mu} F_{\rho)\nu}\, \psi_\nu
	~~+~~
	D_{e}^{\mu\nu\rho} \cdot
	\ov{\psi}{}_e\, \gamma_{(\mu} F_{\rho)\nu}\, \psi_e
	~.
\end{eqnarray}
	As one can see, the absence of strong interactions for leptons
    makes  \eqref{SM_lepton} more compact compared to  
	\eqref{SM_quark}.

\subsection{Higgs sector}	
	The scalar sector of the SM in its minimal form contains one electroweak doublet,
which also admits LV extensions. All LV operators with the use of the Higgs field can 
be further subdivided into two groups. 
	The first are interactions built of the Higgs field and 
	derivatives:
\begin{eqnarray}
\nonumber
\lefteqn{
	\mathcal{L}_{\rm SM}^{\rm Higgs-gauge} =
	}
	\\
\nonumber
	&&
	\phantom{+~~}
	l^\mu \cdot
	i\, 
	H^\dag H \cdot H^\dag \md_\mu H
	~~+~~
	\kappa^{\mu\nu\rho} \cdot
	i\, 
	H^\dag \md_{(\mu} \md_\nu \md_{\rho)} H
	~~+
	\\
\label{Higgs_gauge}
	&&
	+~~
	m^\mu_1 \cdot
	i\, 
	H^\dag F_{\mu\lambda} \md^\lambda H
	~~+~~
	m^\mu_2 \cdot
	i\, 
	H^\dag W_{\mu\lambda} \md^\lambda H
	~~+~~
	\text{h.c.}
	~~+
	\\
\nonumber
	&&
	+~~
	\wt{m}^\mu_1 \cdot
	i\, 
	H^\dag \wt{F}_{\mu\lambda} \md^\lambda H
	~~+~~
	\wt{m}^\mu_2 \cdot
	i\, 
	H^\dag \wt{W}_{\mu\lambda} \md^\lambda H
	~~+
	\\
\nonumber
	&&
	+~~
	n^{\mu\nu\rho}_{1} \cdot
	i\, 
	H^\dag F_{\nu(\mu} \md_{\rho)} H
	~~+~~
	n^{\mu\nu\rho}_{2} \cdot
	i\, 
	H^\dag W_{\nu(\mu} \md_{\rho)}  H
	~~+~~
	\text{h.c.}
\end{eqnarray}

	The second group contains all possible LV extensions of interaction of
	Higgs field and fermions. 
	This group is somewhat larger and includes higher-rank structures.
	The following are the operators involving quarks:
\begin{align}
\notag
	\mathcal{L}_{\rm SM}^{\rm Higgs-quark} & ~~=~~
	h^{\mu}_{QQ} \cdot
	\ov{Q}H \, \gamma_\mu\, H^\dag Q ~~+~~
	\\
\notag
	&
	+~~
	p_{QQ}^\mu \cdot
	\ov{Q}\, \gamma_\mu Q \cdot H^\dag H ~~+~~
	p_{uu}^\mu \cdot
	\ov{u}\, \gamma_\mu u \cdot H^\dag H ~~+~~
	p_{dd}^\mu \cdot
	\ov{d}\, \gamma_\mu d \cdot H^\dag H 
	\\
\label{Higgs_quark}
	&
	+~~
	q^{(1)\mu}_{Qd} \cdot
	\ov{Q}\, d \; \md_\mu H ~~+~~
	q^{(1)\mu}_{Qu} \cdot
	\ov{Q}\, u \; \md_\mu \epsilon H^* ~~+~~
	\text{h.c.}
	\\
\notag
	&
	+~~
	q^{(2)\nu}_{Qd} \cdot
	\ov{Q}\, \sigma^{\mu\nu} d\; \md_\nu H ~~+~~
	q^{(2)\nu}_{Qu} \cdot
	\ov{Q}\, \sigma^{\mu\nu} u\; \md_\nu \epsilon H^* ~~+~~
	\text{h.c.}
	\\
\notag
	&
	+~~
	r^{(1)\mu\nu\rho}_{Qd} \cdot
	\ov{Q}\, \md_{(\mu} \sigma_{\nu)\rho} d \cdot H ~~+~~
	r^{(2)\mu\nu\rho}_{Qd} \cdot
	\ov{Q}\, \sigma_{\nu)\rho} d \; \md_{(\mu} H ~~+~~
	\text{h.c.}
	\\
\notag
	&
	+~~
	r^{(1)\mu\nu\rho}_{Qu} \cdot
	\ov{Q}\, \md_{(\mu} \sigma_{\nu)\rho} u \cdot \epsilon H^* ~~+~~
	r^{(2)\mu\nu\rho}_{Qu} \cdot
	\ov{Q}\, \sigma_{\nu)\rho} u \; \md_{(\mu} \epsilon H^* ~~+~~
	\text{h.c.}
\end{align}
	where $ \epsilon H^* $ is the charge conjugate of the Higgs field.	
	One also has the similar set of operators for interaction of the Higgs 
	with leptons:
\begin{align}
\notag
	\mathcal{L}_{\rm SM}^{\rm Higgs-lepton} & ~~=~~
	h^{\mu}_{LL} \cdot
	\ov{L}H \, \gamma_\mu\, H^\dag L ~~+~~
	p_{LL}^\mu \cdot
	\ov{L}\, \gamma_\mu L \cdot H^\dag H ~~+~~
	p_{ee}^\mu \cdot
	\ov{e}\, \gamma_\mu e \cdot H^\dag H 
	\\
\label{Higgs_lepton}
	&
	+~~
	q^{(1)\mu}_{Le} \cdot
	\ov{L}\, e \; \md_\mu H ~~+~~
	q^{(2)\nu}_{Le} \cdot
	\ov{L}\, \sigma^{\mu\nu} e\; \md_\nu H ~~+~~
	\text{h.c.}
	\\
\notag
	&
	+~~
	r^{(1)\mu\nu\rho}_{Le} \cdot
	\ov{L}\, \md_{(\mu} \sigma_{\nu)\rho} e \cdot H ~~+~~
	r^{(2)\mu\nu\rho}_{Le} \cdot
	\ov{L}\, \sigma_{\nu)\rho} e \; \md_{(\mu} H ~~+~~
	\text{h.c.}
	\\
\notag
	&
	+~~
	\varsigma^{\mu\nu} \cdot
	\left( H^\dag L \right)^T \sigma_{\mu\nu} \left( H^\dag L \right)
	~~+~~
	\text{h.c.}
\end{align}
	The last term in the Higgs-lepton sector, which couples to 
	matrix $ \varsigma^{\mu\nu} $ antisymmetric in the flavor space, is unusual.
	It is special in a sense that it does not have analogues in other sectors, 
	or in lower dimensions as it violates the lepton number by two units.

	This completes the list of the dimension V Lorentz-violating
	operators in the Standard Model.
	The set of operators in the Standard Model appears to be much
	wider than that in QED due to the diversity of fields and interactions.
	Loop corrections are expected to intermix the operators in an 
	even more complicated way. Clearly, there are many operators that give rise 
	to dimension 3 LV interaction with quadratically divergent coefficients. 
	Although, as indicated earlier, studying renormalization of interactions
	is beneficial for refining constraints on LV,
	we are not setting the goal to derive all one-loop RG
	equations similarly to what we have done in QED.

	On the other hand, of particular interest are
	the rank three absolutely
	symmetric operators $ C_X^{\mu\nu\rho} $ 
	and $ \kappa^{\mu\nu\rho} $ that modify the dispersion relations for the SM particles.
	For these operators, we calculate the one-loop RG equations and present 
the results in Appendix \ref{RG_SM}.

%
%
\section{Overview of LV phenomenology at dimension 5 level}
\label{phenomenology}

	We now discuss typical limits on sensitivity to LV dimension 5 operators, 
	which can be inferred from experimental tests 
	of Lorentz symmetry in laboratory, astrophysical observations and
	data on neutrino oscillations.
	Given the abundance of non-minimal interactions we have derived in 
	the last section, it would be useful to separate them in several classes 
        and deduce a typical experimental sensitivity within each class.
	We note that the constraints given in this section only apply to some of the
	components of Lorentz tensors, and the limits on the rest of the components
	typically can be deduced by taking into account the motion of the Earth
	(see {\it e.g.} \cite{clock2}).

We should caution the reader that a detailed analysis of all
phenomenological consequences following from dimension 5 LV operators appears to
be a very complicated task. 
It is so mainly due to such a great variety of LV operators, even if 
one regroups them in several categories.
Given the flavour structures and different Lorentz components one could
easily see that the count of independent components of LV tensors
reaches thousands rather than hundreds. 
This number is far greater than the number of very precise tests of LV. 
It is then apparent that a large number of linear combinations of LV 
parameters may escape prohibitive constraints. 
In this section, we will disregard these subtleties.
Strictly speaking, all limits deduced in this section should be viewed as ``limits
of sensitivity'' to LV, or, in other words these limits tell us what size of
LV observables one should expect, should all {\em e.g.} flavour components
be roughly of the same order of magnitude. 
Nevertheless, our analysis presents a consistent scheme along which one could
proceed to obtain more precise limits on LV parameters, should a more
predictive UV LV theory be found.
	
	Many of the constraints result from laboratory experiments
	or astrophysical observations at energies much lower
	than the weak scale. 
	The Higgs boson, $W$ and $Z$  bosons and heavy SM fermions 
    do not propagate 
	at these energies and can be integrated out.  Such integration at tree level 
    provides new operators of 
	higher mass dimensions which we are not considering here.
	One should keep in mind, however, that loop effects
	admix LV operators with heavy particles to the light quark and lepton 
    LV operators of the same dimension. Such (typically one-loop) corrections 
    include the logarithmic mixing under the RG running,
    as well as the finite threshold corrections. 
	Therefore, the bounds discussed below contain an intrinsic sensitivity to LV 
	interactions involving Higgs and weak bosons.

	For most phenomenological applications it 
 is useful to rewrite the LV  Lagrangian at the normalization scale of
	around $ 1~\GeV $, the borderline of applicability for the quark-gluon description.
	At this scale it is useful to abandon chiral fermions and combine the left- and right-handed
fields into
	full Dirac spinors as well as split the
	$ \sul $ doublets. 
	
	For practical reasons, one can also pass to the mass basis of the 
	flavor matrices, as it facilitates the decoupling of heavy quarks:
\begin{align}
\label{gauge_mass_basis}
\notag
	c_Q^\mu,\  c_u^\mu  & ~~\rightarrow~~ c_{u\phantom{,5}}^\dag 
				\Bigl|_{\rm below\ EW}  ~=~ \frac{1}{2} 
			\left\lgroup W^\dag_u\, c_u^\mu\, W_u ~+~ U_u^\dag\, c_Q^\mu\, U_u 
						\right\rgroup
	\\
	&
	~~\rightarrow~~ c_{u,5}^\mu 
			\Bigl|_{\rm below\ EW}  ~=~ \frac{1}{2} 
			\left\lgroup W^\mu_u\, c_u^\mu\, W_u ~-~ U_u^\dag\, c_Q^\mu\, U_u 
						\right\rgroup
	\\
\notag
	& \dots\dots
	\\
\notag
	& 
	\hspace{-1.0cm}
	u_L ~~\rightarrow~~ U_u\, u_L\,,   \qquad u_R ~~\rightarrow~~ W_u\, u_R\,, \qquad 
	d_L ~~\rightarrow~~ U_d\, d_L\,,   \qquad d_R ~~\rightarrow~~ W_d\, d_R~.
\end{align}


	For consistency of the effective theory, we need to ensure that the operators that we have
	introduced do not transmute into lower dimensions, and thereby not develop quadratic
	divergencies. 
	We can formulate certain criteria to ensure that operators cannot induce lower dimensional
	interactions:
\begin{itemize}
\item	{\it Tensor structure}. Since in the Standard Model there are no $CPT$-odd  
	dimension three operators of rank higher than one, any LV structure that is coupled
	to an irreducible tensor (which is not a vector) is unconditionally protected from
	developing quadratic divergencies.
\item	{\it Supersymmetry}. In the supersymmetric Standard Model, dimension three LV operators
	do not exist at all. 
	Therefore, as long as the theory is considered above the supersymmetry breaking scale, 
	those operators which fall into supermultiplets of the LV MSSM, are protected
\cite{Bolokhov:2005cj}.
	By cancellation of loop contributions due to superpartners, the quadratic divergencies
	turn into logarithmic ones if supersymmetry is exact. 
	It turns out that there is only one type of such operators
\begin{equation}
\label{SUSY_prot}
	\mathcal{L}_{\rm SUSY} ~~=~~
	\wt{c}_{\rm SUSY,Q}^\mu \cdot 
	\left( Y_Q g'\, \ov{Q} \gamma^\lambda \wt{F}_{\mu\lambda} Q   ~~+~~
	g\,  \ov{Q} \gamma^\lambda \wt{W}_{\mu\lambda} Q   ~~+~~
	g_3\, \ov{Q} \gamma^\lambda \wt{G}_{\mu\lambda} Q 
	\right)  ~~+~~
	\ldots~,
\end{equation}	
	(here, $ Y_Q $ refers to the hypercharge of the left quark doublet)
	which, in the case of quarks, must form a certain linear combination to be part of 
	a supermultiplet.
	Linear combinations orthogonal to the above one are not supersymmetric and therefore
	not protected.	
	When the supersymmetry is broken, the above operators are allowed to induce 
	quadratic divergencies, which will be stabilized at the supersymmetry breaking scale. 

\item	{\it $T$-invariance}. Since in the Standard Model one needs multiple loops to flip
	$ T $-parity of flavor-diagonal interactions, one can conclude 
    that the operators which do not have dimension
	three counterpartners with the same $ T $-parity, are protected.

\item	{\it Lepton-number violation}. There are no dimension three LV operators compatible with
	the Standard Model which would violate the lepton number. 
	We know that there is only one $ \Delta L = 2 $ operator of dimension five --- 
	$ \varsigma^{\mu\nu} $, which therefore is protected against developing quadratic divergencies.
\end{itemize}
	
	The operators for which the above criteria do not apply, have no reason to be protected,
	and therefore will intermix with lower-dimensional interactions in a UV-sensitive way. 
	We call such operators ``unprotected''. 
%
	Such interactions are dangerous, and we need to identify them and exclude them 
	from our low energy effective theory. 
	Using $T$-parity it is easy to show that the dangerous operators
	in the quark and lepton sectors (Eqs. \eqref{SM_quark} and \eqref{SM_lepton}) 
	are the ones coupled to the dual field strengths
\begin{align}
\label{SM_divgt}
\notag
	\mathcal{L}_{\rm SM}^{\rm divgt} & ~~=~~
	\wt{c}_{Q,1}^\mu \cdot
	\ov{Q}\, \gamma^\lambda \wt{F}_{\mu\lambda}\, Q 
	~~+~~
	\wt{c}_{Q,3}^\mu \cdot
	\ov{Q}\, \gamma^\lambda \wt{G}_{\mu\lambda}\, Q 
	~~+~~
	\wt{c}_{q,1}^\mu \cdot
	\ov{q}\, \gamma^\lambda \wt{F}_{\mu\lambda}\, q
	~~+~~ 
	\wt{c}_{q,3}^\mu \cdot
	\ov{q}\, \gamma^\lambda \wt{G}_{\mu\lambda}\, q
	\\
	&
	~~+~~ 
	\wt{c}_{L,1}^\mu \cdot
	\ov{L}\, \gamma^{\lambda} \wt{F}_{\mu\lambda}\, L
	~~+~~
	\wt{c}_{e}^\mu \cdot
	\ov{e}\, \gamma^{\lambda} \wt{F}_{\mu\lambda}\, e
	~~+~~
	\mathcal{L}_{\rm Higgs}^{\rm divgt}~,
\end{align}
	where we have abbreviated $ q = u, d $.
	In the Higgs sector, the following operators in 
	Eqs.~\eqref{Higgs_gauge}-\eqref{Higgs_lepton} are unprotected from 
	transmuting into lower dimensional terms:
\begin{align}
\notag
	\mathcal{L}_{\rm Higgs}^{\rm divgt} ~~&=~~
	l^\mu \cdot
	i\, 
	H^\dag H \cdot H^\dag \md_\mu H
	~~+~~
	\wt{m}^\mu_1 \cdot
	i\, 
	H^\dag \wt{F}_{\mu\lambda} \md^\lambda H
	~~+~~
	\wt{m}^\mu_2 \cdot
	i\, 
	H^\dag \wt{W}_{\mu\lambda} \md^\lambda H
	~~+~~
\\
\notag
	& 
	+~~
	h^{\mu}_{QQ} \cdot
	\ov{Q}H \, \gamma_\mu\, H^\dag Q ~~+~~
	h^{\mu}_{LL} \cdot
	\ov{L}H \, \gamma_\mu\, H^\dag L ~~+~~
\\
\label{Higgs_divgt}
	&
	+~~
	p_{QQ}^\mu \cdot
	\ov{Q}\, \gamma_\mu Q \cdot H^\dag H ~~+~~
	p_{uu}^\mu \cdot
	\ov{u}\, \gamma_\mu u \cdot H^\dag H ~~+~~
	p_{dd}^\mu \cdot
	\ov{d}\, \gamma_\mu d \cdot H^\dag H 
	~~+~~
	\\
\notag
	&
	+~~
	p_{LL}^\mu \cdot
	\ov{L}\, \gamma_\mu L \cdot H^\dag H ~~+~~
	p_{ee}^\mu \cdot
	\ov{e}\, \gamma_\mu e \cdot H^\dag H 
	\\
\notag
	&
	+~~
	q^{(2)\nu}_{Qd} \cdot
	\ov{Q}\, \sigma^{\mu\nu} d\; \md_\nu H ~~+~~
	q^{(2)\nu}_{Qu} \cdot
	\ov{Q}\, \sigma^{\mu\nu} u\; \md_\nu \epsilon H^* ~~+~~
	q^{(2)\nu}_{Le} \cdot
	\ov{L}\, \sigma^{\mu\nu} e\; \md_\nu H ~~+~~
	\text{h.c.}
\end{align}

Using the quadratic divergence of the loop corrections generated by operators \eqref{SM_divgt},
one can estimate the strength of the naturalness constraints resulting from experimental limits 
on dimension 3 LV terms
\cite{Heckel:1999sy,clock2,Berglund:}:
\begin{equation}
\label{limit_dim3}
	| b^\mu | ~=~ ({\rm loop\ factor})\, 
		\Lambda^2 | \wt{c}^\mu |  ~\lesssim~ 10^{-29}~{\rm GeV}~.
\end{equation}
	Even in the very conservative assumption about the UV cutoff, 
{\it e.g.} $ \Lambda = \Lambda_{\rm weak} $,
	this limit pushes the interactions \eqref{SM_divgt} far beyond 
	direct probe.
	We again note that even though certain linear combinations \eqref{SUSY_prot} might
	be protected by supersymmetry, below the supersymmetry breaking 
	scale they are unprotected and therefore subject to constraints \eqref{limit_dim3}.

	From this moment, we  concentrate only on UV-safe operators
	and
	list the following effective interactions in the quark sector at the scale of $ 1~\GeV $:
\begin{align}
\nonumber
	\mathcal{L}_{\rm SM\ 1~GeV}^{\rm quark} = \hspace{-2.0cm}&
	\\
\nonumber
	&
	\phantom{+~~}
	c_{q}^\mu \cdot
	\ov{q}\, \gamma^\lambda F_{\mu\lambda}\, q 
	~~+~~
	c_{q,5}^\mu \cdot
	\ov{q}\, \gamma^\lambda \gamma^5 F_{\mu\lambda}\, q 
	~~+~~
	C_q^{\mu\nu\rho} \cdot
	\ov{q}\, \gamma_{(\mu} \md_\nu \md_{\rho)}\, q
	~~+~~
	C_{q,5}^{\mu\nu\rho} \cdot
	\ov{q}\, \gamma_{(\mu} \md_\nu \md_{\rho)}\gamma^5\, q
	\\
\label{SM_quark_1GeV}
	&
	+~~
	D_{q}^{\mu\nu\rho} \cdot
	\ov{q}\, \gamma_{(\mu} F_{\rho)\nu}\, q
	~~+~~
	D_{q,5}^{\mu\nu\rho} \cdot
	\ov{q}\, \gamma_{(\mu} F_{\rho)\nu}\gamma^5\, q
	\\
\nonumber
	&
	+~~
	D_{qg}^{\mu\nu\rho} \cdot
	\ov{q}\, \gamma_{(\mu} G_{\rho)\nu}\, q
	~~+~~
	D_{qg,5}^{\mu\nu\rho} \cdot
	\ov{q}\, \gamma_{(\mu} G_{\rho)\nu}\gamma^5\, q
	~,
\end{align}
	and similarly in the lepton sector:
\begin{align}
\nonumber
	\mathcal{L}_{\rm SM\ 1~GeV}^{\rm lepton} & ~~=~~ 
\label{SM_lepton_1GeV}
	C_l^{\mu\nu\rho} \cdot
	\ov{\psi}\, \gamma_{(\mu} \md_\nu \md_{\rho)}\, \psi
	~~+~~
	C_{l,5}^{\mu\nu\rho} \cdot
	\ov{\psi}\, \gamma_{(\mu} \md_\nu \md_{\rho)}\gamma^5\, \psi
	\\
\notag
	&
	+~~
	D_{l}^{\mu\nu\rho} \cdot
	\ov{\psi}\, \gamma_{(\mu} F_{\rho)\nu}\, \psi
	~~+~~
	D_{l,5}^{\mu\nu\rho} \cdot
	\ov{\psi}\, \gamma_{(\mu} F_{\rho)\nu}\gamma^5\, \psi
	~.
\end{align}

	Passing to the gauge sector, we refer to Eq.~\eqref{SM_gauge}.
	At low energies the LV Lagrangian in the gauge sector
	takes the form
\begin{equation}
\label{SM_gauge_1GeV}
	\mathcal{L}_{\rm SM\ 1~GeV}^{\rm gauge} ~~=~~
	C^{\mu\nu\rho}_\el \cdot 
	F_{\mu\lambda}\, \partial_\nu\, \wt{F}_\rho^{\phantom{\rho}\lambda}
	~~+~~
	C_{\suc}^{\mu\nu\rho} \cdot 
	{\rm tr}\,
	G_{\mu\lambda}\, \mathcal{D}_\nu\, \wt{G}_\rho^{\phantom{\rho}\lambda}
	~.
\end{equation}
	Here, the electromagnetic operator
$ C^{\mu\nu\rho}_\el $
	emerges as a linear combination of the LV
	tensors from the $ \ue $ and $ \sul $ sectors:
\[
	C^{\mu\nu\rho}_\el\Bigr|_{M_W} ~=~ 
		C^{\mu\nu\rho}_{\ue}\Bigr|_{M_W} \cos^2 \theta_W ~+~ 
		C^{\mu\nu\rho}_{\sul}\Bigr|_{M_W} \sin^2 \theta_W~.
\]

	In the Higgs sector, in Eqs.~\eqref{Higgs_gauge}-\eqref{Higgs_lepton} many
	``protected'' terms involve a space-time derivative acting on the Higgs field.
	Below the EW scale, where Higgs does not propagate, such terms do not contribute.
	One is left with the following low-energy interactions:
\begin{align}
\notag
	\mathcal{L}_{\rm SM\ 1~GeV}^{\rm Higgs-induced} & ~~=~~
	\frac{v}{\sqrt{2}}\,r_{q}^{\mu\nu\rho} \cdot \ov{q}\, \md_{(\mu} \sigma_{\nu)\rho} q   ~~+~~
	\frac{v}{\sqrt{2}}\,r_{\psi}^{\mu\nu\rho} \cdot \ov{\psi}\, \md_{(\mu} \sigma_{\nu)\rho} \psi 
	~~+~~
	\text{h.c.}~
	\\
	& 
\label{SM_Higgs_1GeV}
	~~+~~
	\frac{v^2}{2}\,\varsigma_\nu^{\mu\nu} \cdot \nu^T \sigma_{\mu\nu} \nu
	~~+~~
	\text{h.c.}~,
\end{align}
	where $ q = u, d $ and $ \psi = e, \nu $.
	The last operator in Eq.~\eqref{SM_Higgs_1GeV} violates the lepton number by 
	two, and in the low-energy theory can only exist for neutrinos.

	The interactions in \eqref{SM_quark_1GeV}-\eqref{SM_gauge_1GeV} 
	can be divided into two groups.
	The first group is formed by the operators which modify dispersion relations
	and grow with energy \cite{MP:}, which we call the {\it UV-enhanced operators}.
	The second group, correspondingly, hosts all other
	structures, which we designate as ``soft'' LV interactions.
	In Table~\ref{constr_table} we list typical experimental constraints on these
	groups of operators. 
	The numbers in this table have an approximate nature.
	The constraints obtained for the light quark or lepton sectors are transferred
	to the heavier flavors by the renormalization group mixing of the operators:
	the $ 1~\GeV $ limits on a linear combination of the original LV operators
	translate into bounds on the individual operators.
	For the soft interactions, the presented limits have an uncertainty sourced by 
	the simple dimensional analysis of the nucleon matrix elements.
	Nevertheless, these numbers show the order of magnitude of the constraints for the
	corresponding groups of operators.
	We now outline the main sources of these constraints.

\begin{table}[tb]
\caption{Typical constraints for dimension five operators. The constraints do not restrict
	all the components of LV vectors and tensors, however, the limits on unrestricted
	components are induced by boosts created by the motion of the Earth and are
	therefore about three orders of magnitude weaker than the numbers shown.
	}
\label{constr_table}
\begin{tabular}{|ccc|}
\hline
	\multicolumn{1}{|c|}{Operators} & 
	\multicolumn{1}{|c|}{~~~~~~Typical constraints~~~~~~} & 
	\multicolumn{1}{|c|}{Source of constraints} \\
\hline
	\multicolumn{3}{|l|}{\quad Unprotected operators} \\
\hline
	$ \wt{c}_{Q,1}^\mu  $
	$ \wt{c}_{Q,3}^\mu  $
	$ \wt{c}_{q,1}^\mu  $
	$ \wt{c}_{q,3}^\mu  $
	$ \wt{c}_{L,1}^\mu  $
	$ \wt{c}_{\psi}^\mu $  &
	$ \ll 10^{-31}~{\rm GeV}^{-1} $ &
	constraints on dim 3 operators
	\\
\hline
	\multicolumn{3}{|l|}{\quad Operators growing with energy (UV-enhanced operators)} \\
\hline
	$ C_{q}^{\mu\nu\rho} $ $ C_{q,5}^{\mu\nu\rho} $
	$ C_l^{\mu\nu\rho} $  $ C_{l,5}^{\mu\nu\rho} $
	$ C^{\mu\nu\rho}_\el $ &
	$ \lesssim 10^{-33-34}~{\rm GeV}^{-1} $
	&
	high energy cosmic rays
	\\
\hline
	\multicolumn{3}{|l|}{\quad Soft LV interactions} \\
\hline
	$ c^\mu_{q,5} $ $ D^{\mu\nu\rho}_{q,5} $
	$ D_{qg}^{\mu\nu\rho} $  $ D_q^{\mu\nu\rho} $  $ r_q^{\mu\nu\rho} $ &
	$ \lesssim 10^{-28-30}~{\rm GeV}^{-1} $     &
	nuclear spin precession \\
\hline
	$ c^\mu_{q,5} $ $ D^{\mu\nu\rho}_{q,5} $ 
	$ D^{\mu\nu\rho}_{qg,5} $ 
	$ c^\mu_{e,5} $ $ D^{\mu\nu\rho}_{e,5} $ &
	$ \lesssim 10^{-25}~e{\rm cm} $          &
	atomic and nuclear EDMs \\
\hline
	\multicolumn{3}{|l|}{\quad $ \Delta L ~=~ 2 $ interaction} \\
\hline
	$ \varsigma_\nu^{\mu\nu} $          &
	$ \lesssim 10^{-23-24}~\GeV^{-1} $       &
	data on neutrino oscillations \\
\hline
\end{tabular}
\end{table}
	
	{\it Ultra-high energy cosmic rays}. 
	The existence of high-energy cosmic rays of energies $ E_{max} \sim 10^{12}~\GeV $ puts
	stringent bounds on UV-enhanced operators in certain sectors of the Standard Model, 
	depending on relative magnitude of LV sources in these sectors
	\cite{Gagnon:2004xh}. 
	Renormalization group equations (see Appendix \ref{RG_SM}) then spread these limits on the 
	other sectors.
	The UV-enhanced operators modify the dispersion relations of the particles, and this would 
	allow the nucleons in the cosmic rays to emit photons or light leptons, and therefore
	efficiently  lose all their energy before reaching the Earth. 
	The fact of observation of high-energy cosmic rays sets typical constraints on UV-enhanced
	LV of the order of $ 10^{-33-34}~\GeV^{-1} $. 
	It would be fair to say that not all ``corners'' of the parameter space of 
UV-enhanced operators are covered by these
	limits.
	For example, sufficiently strong LV in the up quark sector would allow protons in the cosmic
	rays to decay into $ \Delta^{++} $ which could become stable at high energies
	\cite{Gagnon:2004xh}.
	
	{\it Precision experiments}. 
	Astrophysical constraints are not applicable to soft LV interactions, since the latter do not
	modify propagation of particles.
	For this type of interactions the bounds from low-energy precision experiments are in order.
	The strongest limits occur when an operator induces the interaction 
	of nuclear spin with the nuclear electric 
	or chromomagnetic field. 
	One finds that the operators $ D_q^{\mu\nu\rho} $, $ D_{qg}^{\mu\nu\rho} $ 
	and $ r_q^{\mu\nu\rho} $, 
	when averaged over the nucleus give an effective interaction
\[
	\Leff ~~\propto~~ \ov{N} \partial_{(\mu} \sigma_{\nu)\rho} N ~,
\]
	multiplied by a coefficient $ \sim \LQCD $ which can be estimated by a naive dimensional
	analysis of the nucleon matrix element
\cite{Manohar:1983md}.
	For a non-relativistic nucleus this induces the interaction of the nuclear spin with
	the external preferred directions.
	Known limits
\cite{clock2,Berglund:}
	on interaction of nuclear spin with external directions allow one to estimate typical
	constraints on operators
\[
	| D_{qg}^{0ik} |, ~~| D_{qg,5}^{ijk} |  ~~<~~ 10^{-30}~\GeV^{-1},
	\qquad
	| r_q^{0ik} | ~~<~~ 10^{-31}~\GeV^{-1}~.
\]
	These constraints have an uncertainty related to the knowledge of the 
	spin structure functions which arise for the corresponding nucleon matrix elements.
	The limits on the interactions $ D_q^{\mu\nu\rho} $, $ D_{q,5}^{\mu\nu\rho} $ and
	$ c_q^\mu $ are less strong by a factor of $ \alpha $ due to the suppression of the
	nuclear electric field relative to the chromomagnetic field strength.
	In constraining the leptonic operators $ D_l^{\mu\nu\rho} $ and $ r_e^{\mu\nu\rho} $ one
	loses the advantage of using the strong internal nuclear fields, 
	and the corresponding bounds are weakened by the ratio of the characteristic atomic 
	energy scale to the nuclear energy scale $ p_{at}/p_{nucl} \sim \alpha m_e / \LQCD $.


	{\it Electric dipole moments}. 
	The operators $ D_{q,5}^{\mu\nu\rho} $, $ D_{qg,5}^{\mu\nu\rho} $ and
	$ D_{l,5}^{\mu\nu\rho} $, written in terms of low-energy effective Hamiltonian
	possess the signature of Electric Dipole Moment interactions.
	Averaged over the nucleus, the first two induce nuclear EDMs, for which the 
	existing limits can be used to constrain the amount of Lorentz violation
\cite{Bolokhov:2006yx}:
\[
	|c_{q,5}^0|,~~ | D_{q,5}^{i0k} | ~~\lesssim~~ 10^{-12}~\GeV^{-1}~.
\]
	The electron operator $ D_{l,5}^{\mu\nu\rho} $ applied to paramagnetic atoms induces
	the electric dipole moment of the atom, and this way a bound of the similar strength
	is obtained
\cite{Bolokhov:2006yx}.

	{\it Neutrino phenomenology}.
	The operator $ \varsigma^{\mu\nu}_\nu \cdot \nu^T \sigma_{\mu\nu} \nu $ is capable
	of changing the patterns of neutrino oscillations.
	Constraints from reactor and atmospheric neutrino oscillation data can be used 
\cite{Choubey:2003ke}
	to limit various flavor components of $ \varsigma_\nu^{\mu\nu} $. In the best case scenario, 
	the sensitivity of the neutrino oscillation experiments could provide  the limits at the level of 
\[
	| \varsigma_\nu^{\mu\nu} | ~~\lesssim~~ 10^{-23-24}~\GeV^{-1}~.
\]

	We again comment that the constraints on LV operators displayed in this section do not 
	generically restrict all the components of corresponding LV tensors.
	However, a customary argument applies, that the ``unobservable'' parts of the tensors
	could induce the detectable effects due to the Lorentz boost caused by motion of the Earth 
	relative to the Galaxy, and so cannot exceed the observable components by more than $O(10^3)$.

\section{Conclusion}
	We have presented a systematic Lorentz-violating extension of 
	Quantum Electrodynamics and the Standard Model 
	with all possible dimension five operators.
	Quantum Electrodynamics is the simplest phenomenological example
	to consider. 
	At the level of mass dimension five,
	QED admits a plethora of LV interactions,
	parametrized by background vectors and tensors, most of which are concentrated in the matter 
sector.
We found the one-loop logarithmic renormalization group equations for the QED LV operators, and identified
 one operator that has quadratic sensitivity to the UV-cutoff.

	Extending QED to full Standard Model, we list all possible LV
    interactions satisfying the criteria of the effective field theory 
     given in the Introduction.
	Certainly, the wider gauge group and the diversity of field content lead
	to a sufficiently broader set of LV structures than in QED, although the gauge
	sector remains very simple.
	However, one quickly runs into the problem of dimensional transmutation into 
	dimension three operators via quadratic divergencies, which can easily invalidate the 
    distinction between operators of different dimensions.
	One therefore is naturally lead to an additional requirement of 
the absence of uncontrollable divergencies
	in a consistent effective theory. 
	We identify broad classes of LV operators that are protected against dimensional transmutations 
	by a variety of different mechanisms. These mechanisms include the protection by the irreducibility of 
	the LV spurion tensor structures, supersymmetry, $T$-invariance, and the lepton number conservation. 

	The protected operators of the effective low-energy theory can be divided into several
	qualitatively different groups. 
	The first group of interactions directly affects the propagation of particles 
	by introducing a (species-dependent) ``speed-of-light''. 
	We call such operators UV-enhanced, since they introduce corrections to the
	dispersion relations which grow with the energy. 
	The bounds on UV-enhanced operators are well-known to be of the order 
	$ 10^{-33-34}~\GeV^{-1} $ and come from astrophysical observations. 
	Although these bounds are well studied, we calculate the renormalization 
group equations that govern the logarithmic 
	evolution of these operators over the energy scales, which can be helpful in 
	strengthening the bounds on operators involving heavy fields. 

	All the other protected operators present in our effective theory do not grow with 
energies of propagating particles, and we call them soft LV interactions.
	In general strong astrophysical constraints are not applicable to such
	operators. 
	Limits of the order $ 10^{-28-30}~\GeV^{-1} $ on several classes of such operators
    can be deduced from clock comparison types of 
	laboratory experiments, and are still sufficiently tight. 
	Among less constrained type of operators are those that violate lepton number by two units, 
	and $T$-odd operators that are limited only by the electric dipole moment constraints.

The strength of the analysis performed in this paper is in its generality. 
Indeed, if one day a theory of quantum gravity or any other UV-sensitive theory would 
reach the stage of predicting the low-energy LV phenomenology, such predictions 
can be readily compared with the set of operators derived in this work, and in addition
can be tested for consistency with respect to their UV behavior at the loop level. 

\newpage
%
%
\appendix
\section{RG equations for Dimension Five operators which modify dispersion relations}
\label{RG_SM}


	We list one-loop RG equations for LV operators which modify propagation of
	particles, {\it i.e.} those which couple to absolutely symmetric tensors.
	Although the set of such interactions in the Standard Model is not
	diverse, the equations appear to be complicated.
	The notations for the operators are introduced in section 
	\ref{matter_SM}.
	All operators bear three indices $ \mu $, $ \nu $ and $ \rho $ which we
	omit for brevity.

	In what follows, 
	$ Y_X $  are the hypercharges of the corresponding particles;
	$ \lambda_X $ are Yukawa coupling matrices for species  $ X $;
	$ g' $, $ g $ and $ g_3 $ are correspondingly the $ \ue $, $ \sul $ and $ \suc $ gauge
	coupling constants; 
	we introduce 
%
\begin{align}
\notag
	\alpha_1 & =        
			    \frac{5/3\, N_g ~+~ 1/8}
                                        {6\pi^2}  \\
\notag
	\alpha_2 & =    -\, \frac{19 ~-~ 8 N_g}
		                  {48\pi^2}  \\
\notag
	\alpha_3 & =    -\, \frac{5 ~-~4/3\, N_g}
			         {8\pi^2}~,
\end{align}
	which are
	are the gauge wavefunction renormalization coefficients for the Standard 
	Model, where $ N_g = 3 $ is the number of generations;
	we also use the following notations:
\begin{align}
\notag
	& N_W = ~2 \quad\text{(dim fund SU(2))} &
		\CW  & ~=~ \frac{N_W^2 ~-~ 1}
			       {2 N_W}       
	\\
\notag
	& N_S = ~3 \quad\text{(dim fund SU(3))} &
		\CS & ~=~ \frac{N_S^2 ~-~ 1}
			       {2 N_S}
	~.
\end{align}

	Below we present the RG equations for UV-enhanced LV operators above the EW
	symmetry breaking scale. 
	The Wilson coefficients are assumed to be flavor matrices given in the gauge basis.
\begin{align}
\label{RG_above_ferm}
\notag
	\frac{d}{dt}\, 
	C_L
	&
	~~=~~
	\frac{25}{48\pi^2}
	\left(
		g'^2 Y_L^2  ~+~
		g^2 \CW 
	\right)\, C_L 
	~~+~~
	\frac{1}{32\pi^2} 
		\left\{ \lambda_e\lambda_e^\dag\,,\, C_L \right\}
	\\
\notag
	&
	~~+~~
	\frac{5 g'^2}{48\pi^2}  Y_L^2\, C_\ue \cdot \uflavor
	~~+~~
	\frac{5 g^2}{48\pi^2} \CW\, C_\sul \cdot \uflavor
	\\
\notag 
	&
	~~-~~
	\frac{1}{96\pi^2} \lambda_e C_e \lambda_e^\dag 
	~~-~~
	\frac{1}{64\pi^2} \lambda_e \lambda_e^\dag \cdot \kappa
	\\
\notag
	\frac{d}{dt}\, 
	C_Q
	&
	~~=~~
	\frac{25}{48\pi^2}
	\left(
		g'^2 Y_Q^2 ~+~
		g^2 \CW ~+~
		g_3^2 \CS
	\right)\,
	C_Q
	~~+~~
	\frac{1}{32\pi^2} 
	\left\{ \lambda_d \lambda_d^\dag ~+~
		\lambda_u \lambda_u^\dag \,,\,
		C_Q 
	\right\}
	\\
\notag
	&
	~~+~~
	\frac{5 g'^2}{48\pi^2} Y_Q^2\, C_\ue \cdot \uflavor 
	~~+~~ 
	\frac{5 g^2}{48\pi^2} \CW\, C_\sul \cdot \uflavor 
	\\
\notag
	&
	~~+~~
	\frac{5 g_3^2}{48\pi^2} \CS\, C_\suc \cdot \uflavor
	\\
\notag
	&
	~~-~~
	\frac{1}{96\pi^2}
	\left( \lambda_d C_d \lambda_d^\dag ~+~
		\lambda_u C_u \lambda_u^\dag 
	\right)
	~~-~~
	\frac{1}{64\pi^2}
	\left( \lambda_d \lambda_d^\dag ~-~
		\lambda_u \lambda_u^\dag 
	\right) \cdot
	\kappa
	\\
	\frac{d}{dt}\, 
	C_e
	&
	~~=~~
	\frac{25 g'^2}{48\pi^2} Y_e^2\, C_e
	~~+~~
	\frac{1}{16\pi^2} 
	\left\{ \lambda_e^\dag \lambda_e \,,\,
		C_e 
	\right\}
	\\
\notag
	&
	~~-~~
	\frac{5 g'^2}{48\pi^2} Y_e^2\, C_\ue \cdot \uflavor
	~~-~~
	\frac{1}{48\pi^2} \lambda_e^\dag C_L \lambda_e
	~~+~~
	\frac{1}{32\pi^2} \lambda_e^\dag \lambda_e \cdot
	\kappa
	\\
\notag
	\frac{d}{dt}\, 
	C_u
	&
	~~=~~
	\frac{25}{48\pi^2} 
	\left(
	g'^2 Y_u^2 ~+~ g_3^2 \CS 
	\right)\, C_u
	~~+~~
	\frac{1}{16\pi^2} 
	\left\{ \lambda_u^\dag \lambda_u \,,\, C_u 
	\right\}
	\\
\notag
	&
	~~-~~
	\frac{5 g'^2}{48\pi^2} Y_u^2\, C_\ue \cdot \uflavor
	~~-~~
	\frac{5 g_3^2}{48\pi^2} \CS\, C_\suc \cdot \uflavor 
	\\
\notag
	&
	~~-~~
	\frac{1}{48\pi^2}
	\lambda_u^\dag C_Q \lambda_u 
	~~-~~
	\frac{1}{32\pi^2}
	\lambda_u^\dag \lambda_u \cdot
	\kappa
	\\
\notag
	\frac{d}{dt}\, 
	C_d
	&
	~~=~~
	\frac{25}{48\pi^2} 
	\left( g'^2 Y_d^2 ~+~ g_3^2 \CS 
	\right)\, C_d
	~~+~~
	\frac{1}{16\pi^2} 
	\left\{ \lambda_d^\dag \lambda_d \,,\, C_d 
	\right\}
	\\
\notag
	&
	~~-~~
	\frac{5 g'^2}{48\pi^2} Y_d^2 \, C_\ue \cdot \uflavor
	~~-~~
	\frac{5 g_3^2}{48\pi^2} \CS \, C_\suc \cdot \uflavor
	\\
\notag
	&
	~~-~~
	\frac{1}{48\pi^2} \lambda_d^\dag C_Q \lambda_d 
	~~+~~
	\frac{1}{32\pi^2} \lambda_d^\dag \lambda_d \cdot
	\kappa
\end{align}

	The analogous RG equations for the gauge LV operators take the form:
\begin{align}
\label{RG_above_gauge}
\notag
	\frac{d}{dt}\, 
	C_\ue 
	&
	~~=~~
	- \frac{g'^2}{48\pi^2}\,
	{\rm tr} 
	\lgr
		Y_L^2\, C_L ~+~ N_S Y_Q^2\, C_Q ~-~
		Y_e^2\, C_e ~-~ 
		N_S Y_u^2\, C_u ~-~ N_S Y_d^2\, C_d
	\rgr
	\\
\notag
	&
	\qquad\qquad\qquad\qquad\qquad\qquad\qquad\qquad\qquad
	~~+~~
	\alpha_1 \, g'^2 \, C_\ue
	\\
\notag
	\frac{d}{dt}\, 
	C_\sul 
	&
	~~=~~
	- \frac{g^2}{192\pi^2} {\rm tr} \lgr C_L ~+~ N_S \, C_Q \rgr
	~~+~~
	\left(
		\alpha_2 g^2 ~+~
		\frac{7}{12\pi^2}N_W g^2 
	\right) \cdot
	C_\sul
	\\
	\frac{d}{dt}\, 
	C_\suc
	&
	~~=~~
	- \frac{g_3^2}{192\pi^2} 
	{\rm tr} \lgr
		2 C_Q ~-~ C_u ~-~ C_d 
		\rgr
	~~+~~
	\left(
		\alpha_3 g_3^2 ~+~
		\frac{7}{12\pi^2} N_S g_3^2
	\right) \cdot
	C_\suc
	\\
\notag
	\frac{d}{dt}\, 
	\kappa
	&
	~~=~~
	\frac{5}{12\pi^2} 
	\left[
		(g')^2\, Y_H^2 ~+~ g^2\, \CW
	\right] \cdot \kappa
	\\
\notag
	&
	~~+~~
	\frac{1}{8\pi^2}
	{\rm tr}
	\lgr
		\lambda_e\, \lambda_e^\dag
		~+~
		N_S\, \lambda_u\, \lambda_u^\dag
		~+~
		N_S\, \lambda_d\, \lambda_d^\dag
	\rgr
	\cdot \kappa
	\\
\notag
	&
	~~-~~
	\frac{1}{12\pi^2}
	{\rm tr} 
	\lgr
		N_S\, \lambda_d^\dag\, C_Q\, \lambda_d 
		~+~
		\lambda_e^\dag\, C_L\, \lambda_e
		~-~
		N_S\, \lambda_u^\dag\, C_Q\, \lambda_u
		~-
	\right.
	\\
\notag
	&
	\qquad\qquad\quad
	\left.
		-~
		N_S\, \lambda_d\, C_d\, \lambda_d^\dag
		~-~
		\lambda_e\, C_e\, \lambda_e^\dag
		~+~
		N_S\, \lambda_u\, C_u\, \lambda_u^\dag
	\rgr
	~.
\end{align}
	We observe that mixing of RG operators is quite noticeable between all sectors
	of the Standard Model.

%
%
	The RG equations for UV-enhanced LV interactions below the EW scale read as:
\begin{align}
\label{RG_below_ferm}
\notag
	\frac{d}{dt}\, 
	C_\nu
	&
	~~=~~
	0
	\\
\notag
	\frac{d}{dt}\, 
	C_{\nu,5}
	&
	~~=~~
	0
	\\
\notag
	\frac{d}{dt}\, 
	C_e
	&
	~~=~~
	\frac{25 e^2}{48\pi^2}\,
	C_e 
	\\
\notag
	\frac{d}{dt}\, 
	C_{e,5}
	&
	~~=~~
	\frac{25 e^2}{48\pi^2}\,
	C_{e,5}
	~~+~~
	\frac{5 e^2}{48\pi^2}\, C_\el \cdot \uflavor
	\\
	\frac{d}{dt}\, 
	C_u
	&
	~~=~~
	\frac{25}{48\pi^2}
	\left( q_u^2 e^2 ~+~ \CS g_3^2 \right)\,
	C_u
	\\
\notag
	\frac{d}{dt}\, 
	C_{u,5}
	&
	~~=~~
	\frac{25}{48\pi^2}
	\left( q_u^2 e^2 ~+~ \CS g_3^2 \right)\,
	C_{u,5}
	\\
\notag
	&
	~~+~~
	\frac{5}{48\pi^2} \left( q_u^2 e^2\, C_\el ~+~ g_3^2 \CS\, C_\suc \right)
	\cdot \uflavor
	\\
\notag
	\frac{d}{dt}\, 
	C_d
	&
	~~=~~
	\frac{25}{48\pi^2}
	\left( q_d^2 e^2 ~+~ \CS g_3^2 \right)\,
	C_d
	\\
\notag
	\frac{d}{dt}\, 
	C_{d,5}
	&
	~~=~~
	\frac{25}{48\pi^2}
	\left( q_d^2 e^2 ~+~ \CS g_3^2 \right)\,
	C_{d,5}
	\\
	&
	~~+~~
	\frac{5}{48\pi^2} \left( q_d^2 e^2\, C_\el ~+~ g_3^2 \CS\, C_\suc \right)
	\cdot \uflavor
\end{align}
	for matter operators, and
\begin{align}
\label{RG_below_gauge}
\notag
	\frac{d}{dt}\, 
	C_\el
	&
	~~=~~
	\frac{e^2}{48\pi^2}\,
	{\rm tr}
	\lgr
		C_{e,5} ~+~
		N_S\, q_u^2 \, C_{u,5} ~+~
		N_S\, q_d^2 \, C_{d,5}
	\rgr
	~~+~~
	e^2 \,\alpha_\el \cdot C_\el
	\\
	\frac{d}{dt}\, 
	C_\suc
	&
	~~=~~
	\frac{g_3^2}{96\pi^2}\, {\rm tr} \lgr C_{u,5} ~+~ C_{d,5} \rgr
	~~+~~
	\left(
		\alpha_3  ~+~
		\frac{7}{12\pi^2} N_S 
	\right)g_3^2 \cdot
	C_\suc
\end{align}	
	for gauge LV interactions.
	Here the flavor matrices of the Wilson coefficients are given in the mass basis.
	Below the EW scale we have done an obvious transition to
\[
	\frac{ C_L ~~+~~ C_e }{2} ~~\equiv~~ C_e \Bigr|_{\rm EW}~,
	\qquad
	\frac{ C_L ~~-~~ C_e }{2} ~~\equiv~~ C_{e,5} \Bigr|_{\rm EW}~,
	\qquad
	{\rm etc~.}
\]
	We have denoted the electric charges by $ q_X $, and introduced
\[
	\alpha_\el ~~=~~ \frac{1}{6\pi^2} \times \sum_{\text{species}} q_i^2~,
\]
	which is the wavefunction renormalization coefficient for the electromagnetic field.
	Here the sum runs over all species existing at the given scale $ \mu $.

	Although the RG mixing in Eqs.~\eqref{RG_below_ferm}, \eqref{RG_below_gauge} is not very 
	considerable, we again emphasize that the operators effectively mix above the EW scale, 
	see Eqs.~\eqref{RG_above_ferm}, \eqref{RG_above_gauge}.

\newpage
\section{Young tableaux and irreducible tensors of the Lorentz group}
\label{young}

	To build irreducible tensors of an arbitrary rank one can use
	the Young tableaux. 
	We describe here a recipe how to expand a tensor of a specific
	rank into its irreducible components\footnote{
For more details, the reader is referred to \cite{Hamermesh}.}.
	In the text we most extensively exploit rank three tensors, which
	we here now use as a non-trivial but rather simple example.

	For a tensor of rank $ r $ one builds all possible numbered 
	Young tableaux consisting of $ r $ boxes.
	For each tableau one builds an irreducible component by 
	(anti)symmetrizing its indices as described below.
	After that, to make a component truly irreducible, one has to 
	subtract from it all its $ g^{\mu\nu} $-traces.

	For each numbered diagram, one builds a tensor such that
	each number corresponds to an index ({\it e.g.} 
	for $ T^{\mu\nu\rho} $, one could identify $ 1 \to \mu $,  
	$ 2 \to \nu $,  $ 3 \to \rho $).
	Indices whose numbers form horizontal rows in the diagram
	are symmetrized. 
	Indices which form vertical columns are antisymmetrized.
	Symmetrization always occurs with respect to the name
	of the index.
	Antisymmetrization is always done with respect to the
	position of the index (in this case the number not always
	corresponds to one and the same index).

%
	
	As an illustration to what have been said, we build the diagrams
	for a tensor $ T^{\mu\nu\rho} $.
	One finds four different Young diagrams which can be built out
	of three boxes:
\begin{equation}
\label{rank3_diags}
\raisebox{-0.1cm}{
 \includegraphics[width=1.5cm,height=1.5cm,keepaspectratio]
 {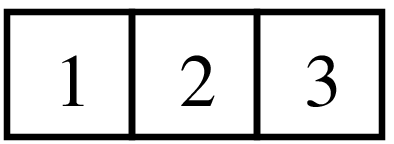}}  
\qquad\qquad
\raisebox{-0.6cm}{
 \includegraphics[width=1.5cm,height=1.5cm,keepaspectratio]
 {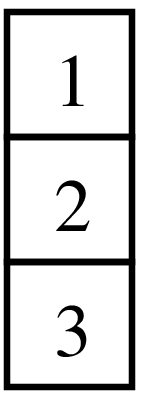}}  
\qquad\qquad
\raisebox{-0.35cm}{
 \includegraphics[width=1.0cm,keepaspectratio]
 {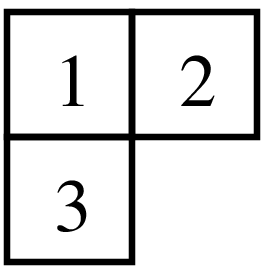}} 
\qquad\qquad
\raisebox{-0.35cm}{
 \includegraphics[width=1.0cm,keepaspectratio]
 {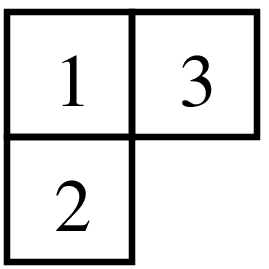}}
\end{equation}

	The first diagram corresponds to an absolutely 
	symmetric component of the tensor:
\[
\raisebox{-0.1cm}{\includegraphics[width=1.5cm,height=1.5cm,keepaspectratio]
 {d3_s.eps}}
	~~\longrightarrow~~
	S^{\mu\nu\rho} ~=~ T^{(\mu\nu\rho)} ~=~
	T^{\mu\nu\rho} ~+~ 	T^{\nu\rho\mu} ~+~	T^{\rho\mu\nu} 
 ~+~	T^{\mu\rho\nu} ~+~ 	T^{\rho\nu\mu} ~+~	T^{\nu\mu\rho}~.
\]
	The second diagram is the absolutely antisymmetric component:
\[
\raisebox{-0.7cm}{\includegraphics[width=1.5cm,height=1.5cm,keepaspectratio]
 {d3_a.eps}}
	~~\longrightarrow~~
	A^{\mu\nu\rho} ~=~ T^{[\mu\nu\rho]} ~=~
	T^{\mu\nu\rho} ~+~ 	T^{\nu\rho\mu} ~+~	T^{\rho\mu\nu} 
 ~-~	T^{\mu\rho\nu} ~-~ 	T^{\rho\nu\mu} ~-~	T^{\nu\mu\rho}~.
\]
	The two ``corner'' diagrams generate, correspondingly,
\[
\raisebox{-0.4cm}{\includegraphics[width=1.0cm,height=1.0cm,keepaspectratio]
 {d3_t1.eps}}
	~~~~\longrightarrow~~~~
	T_1^{\mu\nu\rho} ~=~
	T^{\mu\nu\rho} ~-~ T^{\rho\nu\mu} ~+~ T^{\nu\mu\rho}
	~-~  T^{\rho\mu\nu}~,
\]
	and
\[
\raisebox{-0.4cm}{\includegraphics[width=1.0cm,height=1.0cm,keepaspectratio]
 {d3_t2.eps}}
	~~~~\longrightarrow~~~~
	T_2^{\mu\nu\rho} ~=~
	T^{\mu\nu\rho} ~-~ T^{\nu\mu\rho} ~+~ T^{\rho\nu\mu}
	~-~  T^{\nu\rho\mu}~.
\]

	All four components of \eqref{rank3_diags} 
	(weighed by appropriate coefficients) sum into
	the original tensor $ T^{\mu\nu\rho} $:
\begin{equation}
\label{list_3rank}
	T^{\mu\nu\rho} ~=~
	\frac{1}{3!}\, \left\lgroup S^{\mu\nu\rho} ~+~
				 A^{\mu\nu\rho} ~+~
				 2 T_1^{\mu\nu\rho} ~+
				 2 T_2^{\mu\nu\rho} \right\rgroup~.
\end{equation}
	
	The last step to perform is subtract from each component
	all traces obtained by contraction of any two indices which
	are not antisymmetrized (contraction of antisymmetrized indices is
	trivial).
	The solution can be sought by means of a tensor of a rank less by two:
\begin{equation}
\label{subtr_traces}
	T_{i\;{\rm (irr)}}^{\mu\nu\rho} ~=~
	T_i^{\mu\nu\rho}  ~-~  a_i^{\rho}g^{\mu\nu} 
			  ~+~  a_i^{\mu}g^{\rho\nu} ~+~ ...~,
\end{equation}
	where $ T_i^{\mu\nu\rho} $ is the $ i $-the component
	obtained from the corresponding Young tableau.
	The trace part in the r.h.s. of \eqref{subtr_traces} should 
	possess the same symmetries as $ T_i^{\mu\nu\rho} $
	so as to promote these symmetries to the l.h.s.
	Contracting any two indices in equation \eqref{subtr_traces} and
	requiring the result to vanish one can obtain the explicit expression
	for the trace $ a_i^{\rho} $.
	For the tensors listed in Eq.~\eqref{list_3rank} one obtains:
\begin{align}
\label{irr_3rank}
\notag
	& S^{\mu\nu\rho}_{\rm (irr)} && ~~=~~
		S^{\mu\nu\rho} ~~-~~ \frac{1}{6} 
		\left\lgroup b^\mu g^{\nu\rho} ~+~ b^\nu g^{\rho\mu} ~+~ b^\rho g^{\mu\nu} \right\rgroup
		~,
		&&
		b^\mu ~=~ S^{\mu\lambda\lambda}~,
	\\
	& A^{\mu\nu\rho}_{\rm (irr)} && ~~=~~ A^{\mu\nu\rho}~, 
	\\
\notag
	& T^{\mu\nu\rho}_{1\;\rm (irr)} && ~~=~~
		T_1^{\mu\nu\rho}   ~~-~~
		\frac{1}{3} \left\lgroup  a_{(1)}^{(\mu} g^{\nu)\rho} ~-~
				    2\, a_{(1)}^\rho g^{\mu\nu} 
			    \right\rgroup~,
		&&
		a_{(1)}^\mu ~=~ T^{\mu\lambda\lambda} ~-~ T^{\lambda\lambda\mu}~,
	\\
\notag
	& T^{\mu\nu\rho}_{2\;\rm (irr)} && ~~=~~
		T_2^{\mu\nu\rho}   ~~-~~
		\frac{1}{3} \left\lgroup  a_{(2)}^{(\mu} g^{\rho)\nu} ~-~
				    2\, a_{(2)}^\nu g^{\mu\rho} 
			    \right\rgroup~,
		&&
		a_{(2)}^\mu ~=~ T^{\mu\lambda\lambda} ~-~ T^{\lambda\mu\lambda}
	~.
\end{align}
	These arguments are easily generalized to tensors of arbitrary ranks.

\bibliographystyle{apsrev}
\bibliography{class}

\end{document}